# Magnetism of CuCr₂*X*₄ (*X*= S and Se) spinels studied with muon spin rotation and relaxation (µSR)

Elaheh Sadrollahi[1,2], F. Jochen Litterst[2], Lilian Prodan[3,4], Vladimir Tsurkan[3,4], and Alois Loidl[3]

[1]Institut für Festkörper- und Materialphysik, Technische Universität Dresden, 01069 Dresden, Germany
[2]Institut für Physik der kondensierten Materie, Technische Universität Braunschweig, 38106 Braunschweig, Germany
[3]Center for Electronic Correlations and Magnetism, Institute of Physics, University of Augsburg, 86135 Augsburg, Germany
[4]Institute of Applied Physics, Moldova State University, MD 2028, Chisinau, Republic of Moldova

**Abstract**

We present muon spin rotation and relaxation (µSR) results for chalcogenide spinels CuCr₂*X*₄ with *X*= S and Se. Both compounds are known as ferromagnetic metals with high Curie temperatures. Our µSR and magnetization data show clear signatures for additional magnetic transitions far below the respective Curie temperatures. They can be related to changes in the Cr valence system from the mixed valence between $Cr^{3+}$ and $Cr^{4+}$ at high temperatures with collinear ferromagnetism to a charge-ordered state at low temperatures with a different ferromagnetic structure. Our results demonstrate that the electronic systems and the related spin structures of both compounds are more complex than assumed so far.

## 1. Introduction

The chalcogenide spinels CuCr₂S₄ and CuCr₂Se₄ have been extensively studied in bulk form in the 1960s–70s and characterized as metallic ferromagnets with high ferromagnetic transition temperatures $T_C$ of approximately 375 K and 440 K [1, 2, 3, 4, 5], respectively. High Curie temperatures and their metallicity make these compounds, especially CuCr₂Se₄, of interest for magneto-optics due to their highly spin-polarized character [4, 6].

There have been long-standing controversial opinions concerning the valences of Cu and Cr. Two models have been discussed over the years. Lotgering and van Stapele [3] suggested an electronic configuration of $Cu^+(Cr^{3+}Cr^{4+})S_4^{2-}$, where the ferromagnetism arises from indirect coupling of the $Cr^{3+}$ with $Cr^{4+}$ spins mediated by the S bonds via 90° superexchange, while the $Cu^+$ ions are non-magnetic. The metallicity of the system is explained in terms of hole conduction in the valence band formed by $S^{2-}$-$p$ and $Cr^{4+}$ $3d$ states. In contrast, Goodenough [7] suggested a configuration of $Cu^{2+}Cr_2^{3+}S_4^{2-}$, where the overall ferromagnetism (or, more correctly, the ferrimagnetism) is due to the contributions of two opposing magnetic moments - one associated with the super-exchange interaction of $Cr^{3+}$ ions through 90° S bonds and the other with a $Cu^{2+}$ delocalized hole. The metallic behavior is due to delocalized $Cu^{2+}$ holes.

Electronic band structure calculations of CuCr₂S₄ and CuCr₂Se₄ revealed scenarios different from both models emphasizing the role of covalent mixing between Cr $3d$ orbitals and the anions $p$ orbitals [8, 9]. X-ray absorption spectroscopy (XAS) and X-ray magnetic circular dichroism spectroscopy (XMCD) studies performed at room temperature [10] gave moments for Cr of 2.9 $\mu_B$, as expected for $Cr^{3+}$ and a tiny Cu magnetic moment opposite to that of Cr, thus supporting the Goodenough model, however, Cu is assigned a 1+ valence.

*Ab-initio* calculation studies on these compounds [11] gave more detailed insight into the effects of hybridization leading to delicate changes in the density of states near the Fermi level. The most prominent effect is that the hybridization induces a spin splitting of the otherwise non-magnetic states of the chalcogens. The non-localized moments associated with S and Se, respectively, and Cu sites are supposed to be aligned antiparallel to the moment related to Cr, supporting the conclusions drawn from neutron diffraction [12] and XMCD [10] results.

Unfortunately, pure single crystals of $CuCr_2S_4$ were not grown until now. Electrical resistivity of polycrystalline samples reveals metallic behavior at low temperatures with a tendency to saturation above about 160 K [13]. An anomalous field dependence of Hall resistivities found below 100 K has been related to an anomalous step observed in magnetization near the same temperature, indicating possibly a change of the spin structure [13]. The experimental situation for $CuCr_2Se_4$ is more favorable due to the easy growth of single crystals. At low temperatures, a $T^{3/2}$ metallic behavior of resistivity turns to saturation around 150 K, similar to $CuCr_2S_4$. At higher temperatures, there is a significant increase in the anomalous Hall coefficient [14].

A more recent study of the electric and magnetic resistivities shows a magneto-electronic phase transition at about 65 K, with different linear regions of the temperature dependence of resistivity for temperatures below 60 K and above 100 K [15]. Both regions are metallic but are proposed to possess different ferromagnetic spin structures, conical and uniaxial, respectively.

A Nuclear Magnetic Resonance (NMR) study on both $CuCr_2S_4$ and $CuCr_2Se_4$ using $^{53}Cr$, $^{63}Cu$, and $^{65}Cu$ was interpreted with a mixed valence $Cr^{3+}$-$Cr^{4+}$ state with metallic behavior for high temperatures [16]. Below 80 K, an onset of charge order between $Cr^{3+}$ and $Cr^{4+}$ was proposed that, even at 2 K, was yet only partial. In addition, the charge between Copper and Chromium is redistributed with the formation of $Cu^{2+}$ that is coupling antiferromagnetically to the ferromagnetic Cr sublattice, thus leading to a ferrimagnetic ground state. A comparison of neutron scattering reflections recorded at 88 K and 4.2 K showed changes that were interpreted with the redistribution of electrons between Cu and Cr [17].

Closely related to $CuCr_2S_4$ and $CuCr_2Se_4$ is the $Fe_{1-x}Cu_xCr_2S_4$ system having the same crystal structure. This system reveals a rich magnetic phase diagram reflecting orbital effects and Jahn-Teller instabilities of $Fe^{2+}$ and/or $Cu^{2+}$. Mössbauer spectroscopy, μSR, together with magnetization measurements, showed that for high iron concentrations, Fe is divalent with the spins coupling antiparallel to those of the ferromagnetic Cr lattice, yielding a ferrimagnet [18, 19]. For higher Cu doping, Fe turns more trivalent which is Jahn-Teller inactive. The magnetic ordering temperatures approach that of pure $CuCr_2S_4$. Also, in this concentration range, additional transitions are found at lower temperatures, suggesting a connection to Jahn-Teller active Cr ions, $Cu^{1+}$ being non-magnetic. Extrapolating the Cu doping dependence of these transitions to x=1, i.e., to $CuCr_2S_4$, one of these transitions should occur around 100 K, and a second one should occur around 50 K or below [18].

In the present study, we performed μSR and magnetization measurements on $CuCr_2S_4$ and $CuCr_2Se_4$. The magnetic anomalies traced from magnetization around 100 K, especially for $CuCr_2S_4$, closely resemble those found in $Cu_xZn_{1-x}Cr_2Se_4$, where a transition from helical to axial conical spin structure is found upon increasing Cu concentration x [20]. This change of magnetic structure also gives a relation to the magnetic phase diagram of the $Fe_{1-x}Cu_xCr_2S_4$ system showing similar transitions with varying Cu doping [18]. $CuCr_2Se_4$ reveals similar behavior yet less expressed. Both systems show additional magnetic anomalies at lower temperatures between about 15-30 K that are proposed to be related to changes in anisotropy. While these observed variations in magnetization are only relatively small, they are reflected by strong changes in the μSR signals. We propose that this is caused by changing contributions from local dipolar fields due to changes in the spin orientation.

## 2. Experimental Details

Polycrystalline samples of $CuCr_2S_4$ and $CuCr_2Se_4$ were prepared by solid-state reactions using high-purity elements taken in the stoichiometric ratios. The samples were finally annealed in a chalcogen atmosphere to reduce possible anion defects [21, 22]. The x-ray powder diffraction of the crushed polycrystalline $CuCr_2S_4$ shows diffraction patterns corresponding to the dominant spinel phase and an impurity phase of $CuCrS_2$ less than 8 %, a known antiferromagnet with $T_N = 37.7$ K [23].

Magnetization data have been measured after field cooling (FC) and zero-field cooling (ZFC) and following warming in the temperature range between 2 K and 400 K utilizing a commercial superconducting quantum interference device magnetometer (Quantum Design MPMS-5).

Muon spin rotation and relaxation (μSR) experiments have been performed at the Swiss Muon Source of Paul Scherrer Institute, Switzerland, using a nearly 100% spin-polarized positive muon ($\mu^+$) beam of the πM3 beamline of the General Purpose Surface-Muon instrument (GPS). Spectra were measured on polycrystalline samples in zero field (ZF) and weak transverse field (wTF), perpendicular to the initial muon spin direction.

For the first series of measurements of $CuCr_2S_4$ from 400 K (above $T_C$) down to 5 K, a closed-cycle refrigerator (CCR) was used. The powder-sample pellet was mounted on a Cu-plate sample holder with Al-coated Mylar tape. In a second series of measurements, we used a liquid Helium flow-cryostat for samples of $CuCr_2S_4$ (1.5 - 250 K) and $CuCr_2Se_4$ (5 – 300 K). In this case, the samples were wrapped in aluminized Mylar foil and positioned between the legs of an ultrapure copper fork. The position of the samples was adjusted to lie within the muon beam spot. Thermal contact was achieved by helium exchange gas. The 'VETO' mode [24] was enabled to suppress the background signal from muons stopped outside the samples. For the wTF experiments, a field $B_{app}$= 5 mT was applied. All spectra were analyzed using the program musrFit [25]. Details of the μSR technique are available in the literature, e.g. [26, 27].

## 3. Magnetization Measurements

Fig. 1 shows the temperature dependence of the magnetization (ZFC and FC) of the sulfide compound $CuCr_2S_4$ measured under an applied magnetic field of 5 mT [Fig. 1(a)] and in fields up to 5 T [Fig. 1(b)]. The temperature dependence of the magnetization reveals several transitions. The sharp increase at $T_C = 376$ K (inflection point of M taken with 5 mT) represents the transition from the paramagnetic to the magnetically ordered phase. The magnetization demonstrates a large irreversibility between the ZFC and FC curves that start to separate below $T_C$ due to the contribution of the domain effects. In both ZFC and FC magnetization data, a lowering of magnetization is observed near $T_2 \cong 88$ K, followed by a smaller kink appearing around $T_1 \cong 30$ K. The latter two transitions change upon applying a different magnetic field and finally diminish under an applied field of 5 T, signaling significant changes in the magnetic structure.

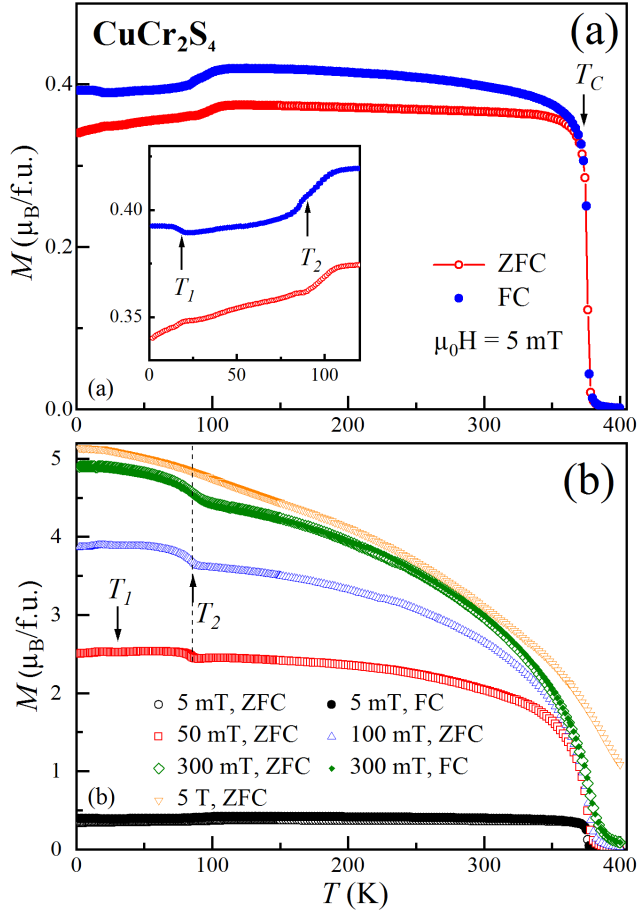

Figure 1: Temperature dependence of the ZFC and FC magnetization of $CuCr_2S_4$ measured in the applied magnetic field of 5 mT (a), and (b) in different applied magnetic fields up to 5 T for.

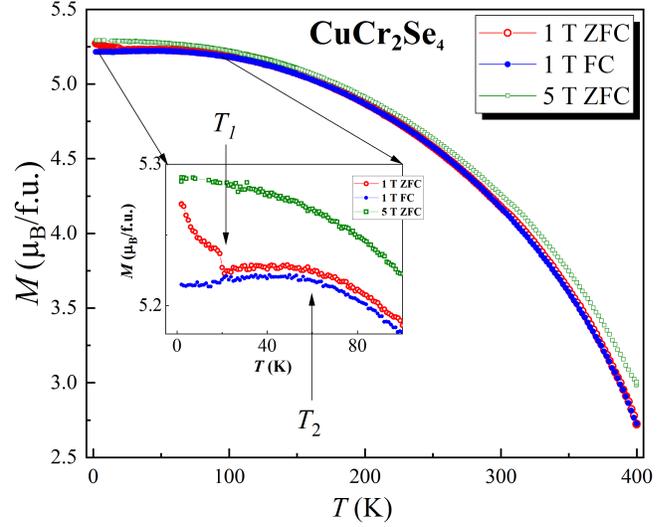

Figure 2: Temperature dependence of the ZFC and FC magnetization measured under an applied field of 1 and 5 T for $CuCr_2Se_4$. The inset shows a magnified range of the low-temperature data.

Fig. 2 presents the temperature dependence of the magnetization of the selenide compound $CuCr_2Se_4$. Note that these data were collected in experiments that were not started from above $T_C$ but only from 400 K. Both ZFC and FC curves measured under 1T reveal a shallow maximum at around 60 K. The ZFC curve at 1 T shows a sudden increase at about 20 K. The FC magnetization curve reveals a smooth saturation down to 2 K. The field dependence of the magnetization at different temperatures in $CuCr_2S_4$ and $CuCr_2Se_4$ is presented in Fig. 10 of the Appendix. The saturation of magnetization is reached in rather low fields, and coercive fields are negligible, as expected for soft ferro(ferri)-magnets.

## 4. Muon Spin Rotation and Relaxation

In μSR measurements, spin-polarized muons are implanted into the material to be studied. Muons stop at interstitial sites, where they precess around the total magnetic field $B_\mu$ acting at the muon site.

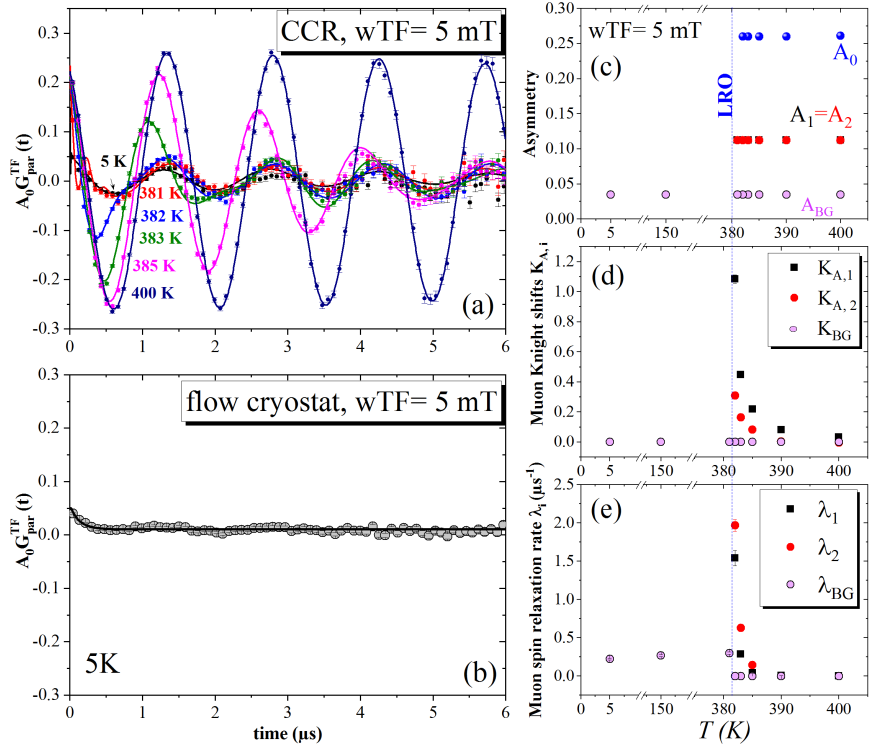

Figure 3: (a) μSR rotation patterns of CuCr$_2$S$_4$ at selected temperatures taken in a weak transverse field (wTF) of 5 mT in CCR cryostat, and (b) rotation pattern at 5 K in flow cryostat. The corresponding fits using eq. 1 are shown in solid lines. A$_0$ is the initial (total) asymmetry and G$_{par}^{TF}$(t) the muon spin polarization function. (c) Temperature dependence of A$_0$ and the asymmetries A$_i$ (A$_1$, A$_2$, A$_{BG}$) obtained from fits to the wTF μSR time spectra using eq. 1. (d) Muon Knight shifts K$_{A,1}$ and K$_{A,2}$ at muon stopping sites 1 and 2. The rotation frequency of the Cu background is nearly constant down to 5 K. Since this frequency serves as reference K$_{BG}$ = 0. (e) wTF relaxation rates (λ$_1$) display clear peaks in the vicinity of the magnetic phase transition T$_C$ and onset of the long-range magnetic order.

In the experiment, the time evolution of the asymmetry A(t) of the decay positron count rates is followed from which one can derive the time-dependent polarization of the muons after implantation at time t=0: A(t)/A(t=0)=G$_z$(t).

In the paramagnetic state, the wTF polarization function will oscillate with a frequency ν$_μ$=γ$_μ$B$_μ$/2π. γ$_μ$=2π×135.5 μs$^{-1}$ T$^{-1}$ is the muon gyromagnetic ratio. Here, B$_μ$ corresponds to the applied field B$_{app}$ (eventually corrected for Knight-shift, see below). Upon onset of magnetic order, there appear additional local-fields B$_μ$ that in a powder specimen will add at random directions to B$_{app}$. Since the local fields, B$_μ$, are significantly stronger than B$_{app}$, there will be severe damping of the wTF rotating signal, leading to an apparent vanishing of this signal. Note also that an applied field may enter a ferromagnetic sample only if it is larger than the saturation field. Thus, for very small applied fields, the wTF rotating signal will also vanish in the ordered state. If the specimen is only partially ordered, a signal will stay left rotating in B$_{app}$ with an asymmetry corresponding to the paramagnetic volume fraction of the sample. The wTF μSR spectra of CuCr$_2$S$_4$ depicted in Fig. 3(a) were collected using the CCR setup. They show a distinct change in spectral shape between 400 K and 381 K. Spectra at 5 K were collected using the CCR and the flow cryostat setups to obtain an overlapping data point for a comparison of data collected with either of the two setups. There is no precessing signal (i.e., there is no background signal) at 5 K when using the flow cryostat (see Fig. 3(b)), but when using the CCR, a precessing signal with an initial asymmetry of 0.035 (14% of initial total asymmetry $A_0 = 0.26$) corresponding to the applied field of 5 mT is observed.

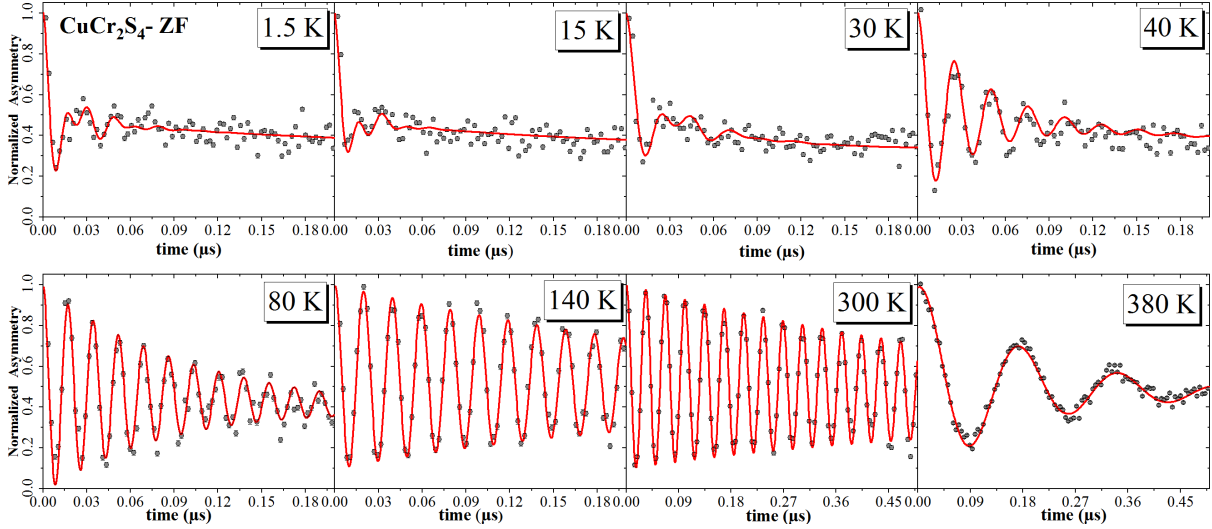

Figure 4: ZF μSR spectra of CuCr$_2$S$_4$ at selected temperatures for a 0 - 0.2 μs time window for 1.5 K - 140 K and 0 - 0.5 μs time window for 300 K - 380 K showing the spontaneous muon spin rotation in the magnetically ordered state. Solid lines are the fits to the data using equations 2 and 3.

This additional asymmetry arises from muons hitting the Cu sample holder, causing a background signal in the CCR setup. Above $T_C \cong$ 381 K, the spectra consist of three sub-signals: one undamped signal due to this background ($A_{BG} \sim$ 14% of $A_0$) oscillating with the frequency corresponding to the applied field. The other two signals correspond to the majority of muons stopping in the sample (see Fig. 3(c)) and are attributed to two muon-stopping sites, 1 and 2. One signal is slightly stronger damped than the other, and their frequencies are strongly temperature dependent (see Fig. 3(d, e)).

We relate this change of frequency above $T_C$ to the muon Knight shift. The applied field $B_{app}$ weakly polarizes the paramagnetic spin ensemble; thus, the magnitude of $B_\mu$ is not precisely that of $B_{app}$.

We analyzed the wTF μSR spectra by a sum of the three contributions:

$$A_0 G_{par}^{TF}(t) = \sum_{i=1}^{2,\,BG} A_i\, e^{-\lambda_i t}\, \cos\left(2\pi\nu_i t + \frac{\pi\varphi}{180}\right) \quad (1)$$

Here, $A_i$ corresponds to the asymmetries of signals from sites 1 and 2, and $A_{BG}$ refers to the background due to the Cu sample holder. Due to the used spin rotator, there is a phase shift of about 30° with $A_0 = 0.26$. Since fits of the asymmetries $A_1$ and $A_2$ are not very sensitive we have fixed the asymmetry ratio $A_1$:$A_2$ to 1:1. $\lambda_i$ are the muon depolarization rates (paramagnetic damping). The obtained temperature dependences of the damping rates and μSR Knight shift $K_{A,i} = (\nu_i - \nu_{BG})/\nu_{BG}$ of the rotation signals, Fig. 3(d), indicate an onset of long-range magnetic order (LRO) at about 380 K, in good agreement with the onset of rising magnetization above $T_C = 373$ K (as assigned from the inflection point of magnetization). As reference frequency for the Knight shift, we have chosen the rotation frequency of the Cu background signal $\nu_{BG}$, which stays nearly constant in the whole temperature range and shows no changes at $T_C$. As already mentioned, the damping rates $\lambda_1$ and $\lambda_2$ at the muon stopping sites

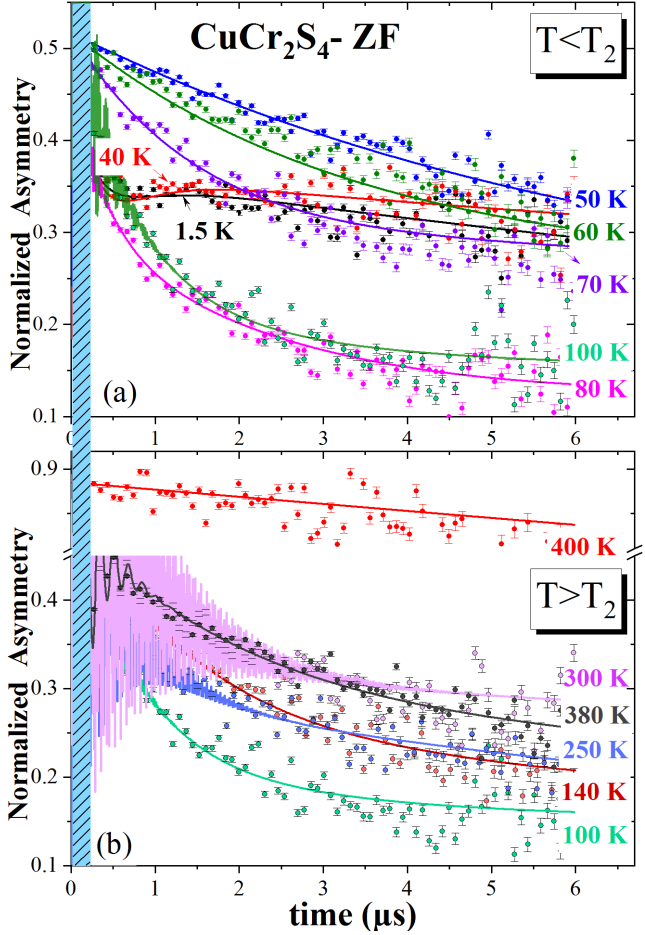

Figure 5: ZF μSR spectra of CuCr$_2$S$_4$ depicted with lower time resolution extending to longer times up to 6 μs showing the variation of dynamic relaxation at selected temperatures T < T$_2$∼ 80 K in (a) and T > T$_2$ in (b). The spectra have been shifted vertically. A shaded blue rectangle is shown in Fig. 4, where a zoomed-in view of the shorter time in the magnetically ordered state. Solid lines are the fits to the data using equations 2 and 3.

1 and 2 are different (see Fig. 3 (e)). One possible interpretation is that the muon spins in the two sites are exposed to different slowing-down processes of the electronic spin system upon approaching T$_C$ from above.

The relaxation rate λ$_1$ is slightly larger than λ$_2$, which means slower paramagnetic spin fluctuations and, thus, stronger correlations between neighboring electronic spins at this site. In addition, we cannot exclude, however, contributions from different inhomogeneous broadening at the two muon-stopping sites caused by a distribution in ordering temperature.

The zero field μSR (ZF) spectra of CuCr$_2$S$_4$ at several temperatures are displayed in Fig. 4 and Fig. 5 within short and long-time windows. Good approaches could be achieved by adding three components (labeled: osc, LKT, BG):

$$A_0 G_{LRO}^{ZF}(t) = A G_{osc}^{ZF}(t) + A_{BG} \qquad (2)$$

for temperatures above T$_1$ up to T$_C$, and

$$A_0 G_{LRO}^{ZF}(t) = A G_{osc}^{ZF}(t) + A_{LKT} G_{LKT}^{ZF}(t) + A_{BG} \qquad (3)$$

for temperatures below T$_1$.

where

$$G_{osc}^{ZF}(t) = \sum_{i=1}^{2} A_i \left[ \frac{2}{3} e^{-\lambda_{T, i} t} \cos\left(2\pi \nu_{\mu, i} t\right) + \frac{1}{3} e^{-\lambda_{L, i} t} \right] \qquad (4)$$

and

$$G_{LKT}^{ZF}(t) = \frac{2}{3} e^{-a_{LKT} t} (1 - a_{LKT} t) + \frac{1}{3} . \qquad (5)$$

$G_{osc}^{ZF}(t)$ represents the sum of the signals with the rotation frequencies $\nu_{\mu, i}$ of the muon spin in the local internal fields caused by ordered electronic magnetic moments (first term in the square bracket of eq. 4). $\lambda_{T, i}$ are the transverse damping parameters of the oscillations mainly caused by field inhomogeneities. The second term in the square bracket, the "1/3 tail", is related to field dynamic fluctuations along the direction of initial muon polarization. $\lambda_{L, i}$ are the longitudinal damping parameters.

For T ≤ 40 K, that is close to T$_1$, an additional time-dependent component with lower asymmetry is necessary for achieving an acceptable fit. This is possible by using a so-called Lorentzian Kubo-Toyabe function $G_{LKT}^{ZF}(t)$. It is the spectral response of random local fields at the muon sites distributed around a mean value of zero. While for a concentrated system of randomly oriented static dipoles, the probability distribution of the internal local field is a Gaussian field distribution, the field distribution in our samples is better described by a Lorentzian distribution as found for systems with dilute dipole moments. The decay rate $a$ of polarization is related to the width of the field distribution [28, 29], $a_{LKT}/γ_μ$ being the half-width at half maximum. Initially, the muon spin polarization decays exponentially, then passes through a minimum, and after recovering to 1/3, remains time-independent.

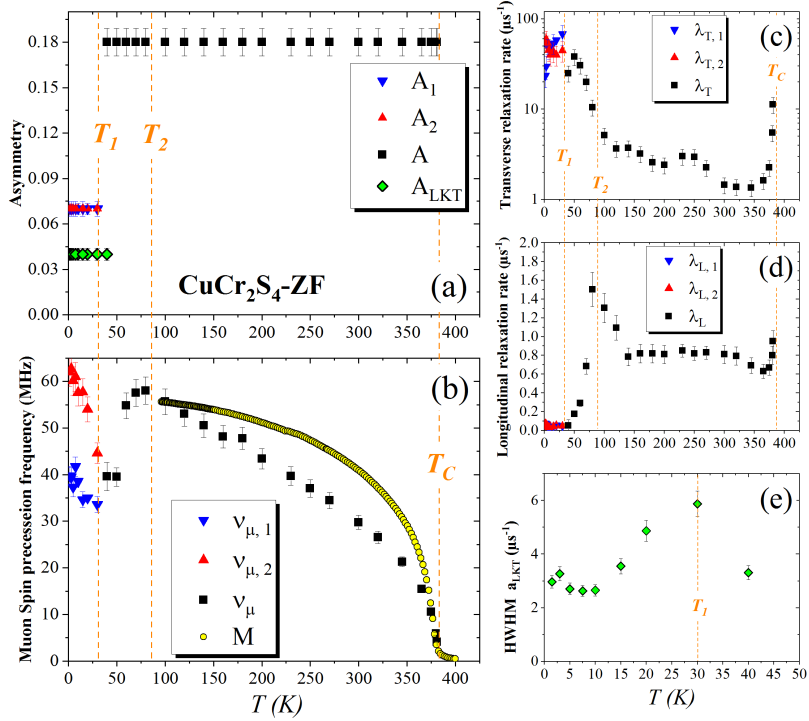

Figure 6: Temperature dependence of ZF μSR parameters in CuCr₂S₄: (a) the asymmetries, A, $A_i$, $A_{LKT}$, (b) spontaneous μ⁺ spin precession frequencies, ν and $ν_{μ,i}$. The open symbols M in (b) are the experimental magnetization data measured in an external field of 1 T normalized to the μSR rotation frequencies at 100 K. (c) transverse relaxation rates $λ_T$ and $λ_{T,i}$, (d) longitudinal relaxation rate $λ_L$ and $λ_{L,i}$, (e) half-width at half maximum, HWHM $a_{LKT}$. Dashed lines indicate the magnetic phase transitions identified in magnetization measurements.

$A_{BG}$ is due to the non-magnetic background (BG) signal coming from the Cu holder in CCR measurements (i.e., this component was only used for the data recorded using CCR, where $A_{BG}$ was fixed at 0.035; see Fig. 3).

The ZF μSR spectra between 5 K and $T_C$ show clear oscillations at the early times (high resolution) that prove a magnetically ordered state (see Fig. 4). We can distinguish three temperature regions. The range between $T_2 \sim 80$ K up to $T_C$ can be well described by the formalism of eq. 2 yet using only one oscillation frequency.

In the temperature region between $T_1 < T < T_2$, the frequency shows a distinct variation accompanied by changes in the damping parameters. Below $T_1$, the ZF spectra are more complex, with severely damped rotation signals having different frequencies that can be well described by the formalism of eq. 3.

The development of the asymmetry parameters for the oscillating and the relaxation parts is shown in Fig. 6(a). The fact that the asymmetry amplitude is shared between two signals (one osc + BG) above $T_1$ and four signals (two osc + LKT+BG) below $T_1$, is evidence that the muons are stopping in different magnetic environments. A, the asymmetry of the single rotation signal above $T_1$, is divided into two rotation signals with equal asymmetries, $A_1$ and $A_2$, and a Lorentzian Kubo-Toyabe signal $A_{LKT}$ below $T_1$.

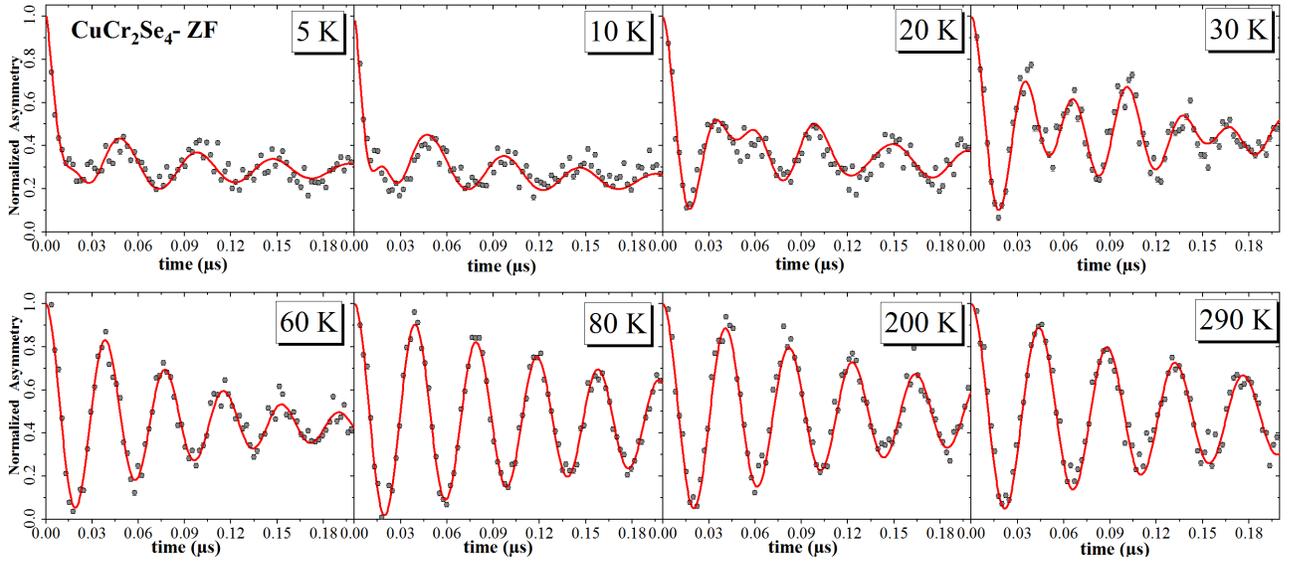

Figure 7: ZF μSR spectra of CuCr₂Se₄ at selected temperatures for a 0-0.2 μs time window showing the spontaneous muon spin rotation in the magnetically ordered state. Solid lines are the fits to the data using equations 2 and 3.

As seen in Fig. 6(b), the temperature dependence of the frequency $\nu_\mu$ follows a magnetization-like curve down to $T_2$. Between $T_2$ and $T_1$, $\nu_\mu$ decreases to about 40 MHz. The two rotation signals detected below $T_1$ have frequencies $\nu_{\mu,1}$ increasing from 40 MHz at $T_1$ up to about 60 MHz at 1.5 K, while $\nu_{\mu,2}$ is increasing from 25 MHz to about 40 MHz at 1.5 K. The change in frequency below $T_2$ is accompanied by a significant increase in transverse and longitudinal relaxation rates, as seen in Fig. 6(c) and 6 (d).

The temperature dependence of the longitudinal relaxation rates ($\lambda_L$ and $\lambda_{L,i}$) and the HWHM $a_{LKT}$ of the Kubo-Toyabe signal are shown in Fig. 6(d) and 6(e), respectively. The values of $\lambda_{L,1}$ and $\lambda_{L,2}$ cannot be determined independently and therefore were assumed to be equal, i.e. the muon spins in the associated sites have the same underlying spin dynamic rates revealing a maximum near $T_2$ and an increase close to $T_C$ where also the transverse damping λ$_T$ is increasing. The value of HWHM a$_{LKT}$ (see Fig. 6(e)) shows a clear peak at $T_1$.

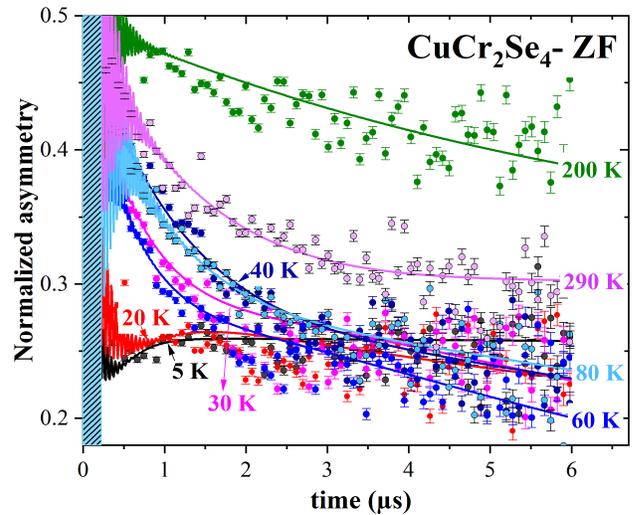

Figure 8: ZF μSR spectra of CuCr₂Se₄ depicted with lower time resolution extending to longer times up to 6 μs showing the variation of dynamic relaxation at selected. The spectra have been shifted vertically. A shaded blue rectangle is shown in Fig. 7, where a zoomed-in view of the shorter time in the magnetically ordered state. Solid lines are the fits to the data using equations 2 and 3.

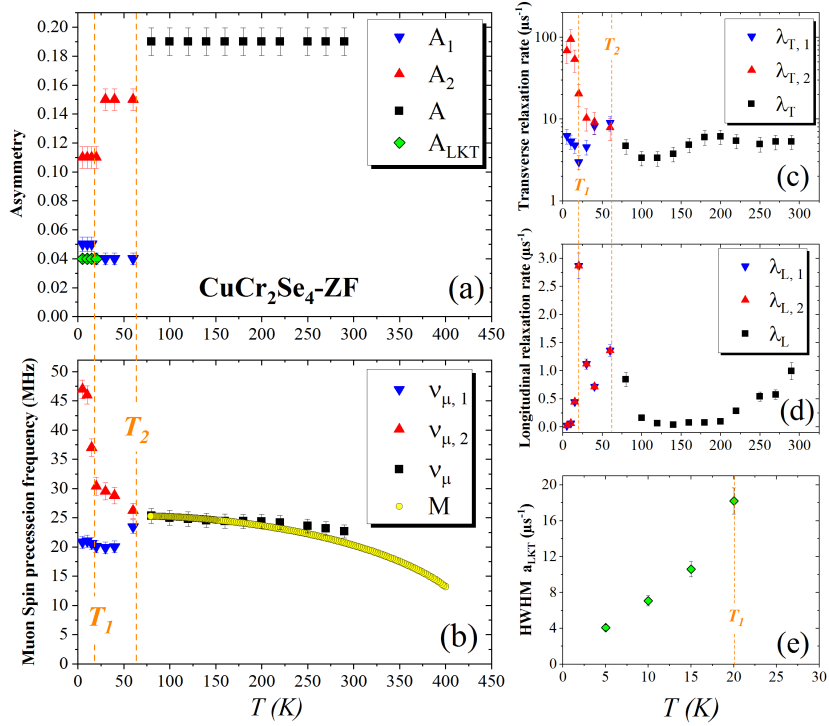

Figure 9: Temperature dependence of ZF μSR parameters in CuCr$_2$Se$_4$: (a) the asymmetries, A, A$_i$, A$_{LKT}$, (b) spontaneous μ$^+$ spin precession frequencies, ν and ν$_{\mu,i}$. The open symbols M in (b) are the experimental magnetization data measured in an external field of 1 T normalized to the μSR rotation frequencies at 80 K. (c) transverse relaxation rates λ$_T$ and λ$_{T,i}$, (d) longitudinal relaxation rate λ$_L$ and λ$_{L,i}$, (e) half-width at half maximum, HWHM a$_{LKT}$. Dashed lines indicate the magnetic phase transitions identified in magnetization measurements.

ZF μSR spectra of CuCr$_2$Se$_4$ could only be taken in the magnetically ordered regime up to 290 K. They are very similar to those of CuCr$_2$S$_4$, with severely damped signals at low temperatures.

The ZF μSR spectra at several temperatures between 5 K and 290 K are displayed in Fig. 7 and Fig. 8 within short and long-time windows. These spectra show clear oscillations at early times (high resolution). Between 290 K and $T_2 \cong 60$ K, the spectra can be reproduced with one rotating signal and can be well-fitted using eq. 2. Below $T_2$, the spectra comprise two rotation signals with severely damped oscillations due to significant changes in transverse and longitudinal relaxation rates, as seen in Fig. 7 and Fig. 8. The temperature dependences of parameters derived using equations 2 and 3 are depicted in Fig. 9. The asymmetries (Fig. 9(a)) show the same tendency as seen for CuCr$_2$S$_4$; the asymmetry of the rotation signal A is temperature independent above $T_2$. Below $T_2$, however, A is shared between two rotation signals with asymmetries A$_1$ and A$_2$ in a ratio of about 1:4. Below $T_1 \cong 20$ K, the ratio of asymmetries A1:A2 changes to about 1:2, while A$_{LKT}$ is nearly constant.

Below $T_2$, we find a clear onset of two precession signals [see Fig. 9(b)] with strongly varying frequencies, especially below $T_1$. These anomalies are reflected also in peaks in transverse and longitudinal relaxation rates at $T_2$ and $T_1$ (see Fig 9(c) and (d)) and also in a$_{LKT}$.

## 5. Discussion

While the ZF μSR spectra below the respective Curie temperatures of $CuCr_2S_4$ and $CuCr_2Se_4$ exhibit spontaneous rotation patterns with frequencies following magnetization-like temperature dependencies, there are found clear deviations from this behavior below about 100 K. Below the temperatures $T_2 \approx 80$ K and 60 K (for $CuCr_2S_4$ and $CuCr_2Se_4$, respectively) at which anomalies in magnetization were found, the muon signals in both studied chalcospinels are distinctly different from those found at higher temperatures. At still lower temperatures, $T_1 \cong 30$ K and 20 K, respectively, there follow further changes that can be related to small anomalies in magnetization, mainly apparent in ZFC and FC curves.

Let us first consider the situation below $T_1$. Notably, $T_1$ for $CuCr_2S_4$ is close to the antiferromagnetic transition of the detected impurity phase $CuCrS_2$ at 37.7 K (as corroborated by our specific heat data). Therefore, we cannot entirely exclude some perturbation of local fields at the muon sites below $T_1$. On the other hand, the spectra of $CuCr_2Se_4$ reveal very similar behavior, yet there are no indications for the presence of an impurity, and we conclude that the changes in local magnetic fields near $T_1$ should be related to changes in spin orientations of the cations.

Changes in locally acting magnetic fields may, in fact, also occur at low temperatures when muons, after their implantation, are initially stopped in a metastable state of the electrostatic potential in their electronic surrounding. The relaxation to a stable ground state may not happen within the muon lifetime of 2.2 μs, which is achieved only for higher temperatures. Though we cannot entirely exclude this scenario, we note that the changes in spontaneous frequencies found at $T_1$ and also $T_2$ for $CuCr_2S_4$ and $CuCr_2Se_4$ are distinctly related to changes in macroscopic magnetization that details of muon implantation would not cause. Also, we cannot trace any changes in the initial phase of the oscillating spectral components that would be expected upon repopulation of muon sites with increasing temperature. What, however, we cannot exclude, is a possible change of muon stopping positions near $T_1$ and $T_2$ due to structural changes related to the magnetic transitions. By now we have yet no indications from XRD of such structural changes. Magnetization anomalies near $T_2$ (the step for $CuCr_2S_4$, the shallow maximum for $CuCr_2Se_4$), as shown in Figures 1 and 2, have also been reported earlier in reference [13].

The shape of magnetization curves below $T_2$, especially for $CuCr_2S_4$, is very similar to those reported for $Cu_xZn_{1-x}Cr_2Se_4$ at high Cu concentrations [20]. While $ZnCr_2Se_4$ is known as helimagnet, the admixture of Cu leads gradually via a conical spin arrangement for intermediate compositions to an axial ferromagnet for $CuCr_2Se_4$. Notably, $T_2$ is close to the temperatures for which changes in electrical resistivity and the Hall effect indicate a change in electronic transport related to the formation of intermediate valence of $Cr^{3+}$-$Cr^{4+}$ pairs above $T_2$.

Thus, it appears plausible to interpret the observed changes in μSR rotation frequencies, transverse and longitudinal damping, near $T_2$ with changes in the local magnetic fields and their dynamics at the muon site related to magnetic transitions between axial ferromagnetism at high temperatures and conical (or more complicated) structures below $T_2$. In fact, the changes observed in rotation frequencies $\nu_\mu$ are by far stronger than the variations of magnetization. The reason can be sought in the contributions to the local field at the muon position by contact fields due to polarized conduction electrons and dipolar fields due to localized spins from neighboring cations. While the contact fields are in first approximation isotropic, the dipolar fields will vary distinctly upon changes in the spin structure for the proposed change from a collinear ferromagnetic to a conical structure and may cause changes in $\nu_\mu$ below $T_2$. A similar behavior of $\nu_\mu$ has been observed in ferromagnetic gadolinium metal ($T_C = 293$ K) [19, 27, 30, 31]. The precession frequency $\nu_\mu$ has a peculiar temperature dependence due to the turning away of the magnetization from the $c$-axis beginning at 245 K [19]. In our presently studied compounds, this rotation of electronic spins is reflected in a lowering of $B_\mu$ (derived from $2\pi\nu_\mu = \gamma_\mu B_\mu$) compared to the values expected from the magnetization curve. In parallel with the reduction of $B_\mu$ at about $T_2$, we find a strong rise in transverse relaxation rates $\lambda_T$ within the spin-reoriented phase, indicating that the magnetic moments do not turn uniformly. The observed maxima in longitudinal relaxation rates $\lambda_L$, are indicative for magnetic transitions.

For both compounds, are found distinct changes in the ZF spectra below the respective temperatures $T_1$. Firstly, a second spontaneous frequency is observed. The longitudinal relaxation rates approach zero for the lowest temperatures as expected in the static limit.

In addition, a contribution with constant asymmetry amounting to about 20% of total asymmetry and a shape that can be pragmatically approximated by a Kubo-Toyabe function is observed. This signal shows no defined muon spin rotation signal but represents a static random field distribution with (assumed) Lorentzian shape around zero field. There is currently no theoretical justification for the use of a Lorentzian Kubo-Toyabe function as part of the signal from a magnetically ordered sample. Equation 3 is used primarily to quantify the data and the inclusion of the Lorentzian Kubo-Toyabe does not affect the main results. We want however point out that a possible origin of the Lorentzian Kubo-Toyabe signal may be muons in sites where the different local field sources (mainly dipolar) are mutually cancelling. The appearance of this additional signal points to a change in muon stopping positions below $T_1$, possibly caused by a structural change, as indicated earlier. The Lorentzian-shaped field distribution is only caused by distant magnetic inhomogeneities. Muon sites in highly symmetric surrounding sensing (nearly) zero local field and a small Lorentzian field distribution have, e.g., been reported in cubic magnets like uranium monopnictides [32, 33].

In $CuCr_2S_4$, the onset of long-range magnetic order at $T_C \approx 380$ K is in good agreement with magnetization data. The magnitude of the mean magnetic field $B_\mu$ at the muon sites should then follow roughly the magnetization curve. However, as depicted in Fig. 6 (b), the magnetic field $B_\mu$ shows a remarkably steeper temperature dependence compared to the magnetization curve. A special feature is the nearly linear dependence of $B_\mu$ between 300 K and 100 K, deviating from the significantly curved dependence of the $M(T)$. Due to the complex superposition of several in part counteracting magnetic fields contributing to $B_\mu$ in a ferro- (or ferri-) magnetic material, we desist here from any attempt of a quantitative interpretation [34].

For temperatures between $T_2$ and 290 K, the μSR spontaneous rotation frequencies of $CuCr_2Se_4$ show only little variation in agreement with the small decrease of magnetization in this temperature range, which contrasts the situation in $CuCr_2S_4$. Also, the rotation frequencies are distinctly lower, indicating that the relative contributions from contact and dipolar fields are different from those in $CuCr_2S_4$. This can be expected, since both materials (but especially $CuCr_2S_4$) cannot be counted as "good metals".

Evidence for different muon stopping sites also comes from the TF μSR spectra of $CuCr_2S_4$ near $T_C$ that can be fitted to two signals with different dampings and different Knight shifts. We can only speculate about stopping sites in these materials since proper density functional theory (DFT) calculations are missing. Possible candidates are sites located close to the chalcogen ions, e.g., as found in the $A$-site spinel $CuAl_2O_4$ [35]. Muons sensing different spin correlations in different stopping sites are, e.g., found in correlated paramagnets like in $A$- site spinels $ACr_2O_4$ ($A$ = Fe, Mn, Co) [36], thio-spinels $FeSc_2S_4$ [37], or $CuCrZrS_4$ [38]. It was suggested that in these systems, the coexistence of differently strong spin-correlated volume fractions is related to the presence of frustration by competing exchange interactions. In the present case, however, we have yet no definite picture of its origin. It may, e.g., be related to cation disorder.

Transitions in the temperature range of $T_2$ and $T_1$, as found in $CuCr_2S_4$ and $CuCr_2Se_4$, could be expected from an extrapolation of the magnetic transitions observed in the related system $Fe_{1-x}Cu_xCr_2S_4$ with varying Cu substitution levels [18] (a detailed discussion will be given elsewhere [39]). Supposed origins of these transitions are Jahn-Teller active ions. The dominant species of iron for high Cu concentrations is $Fe^{3+}$, which is not Jahn-Teller active and, therefore, cannot cause the changes found in magnetic behavior. One Jahn-Teller active ion could be $Cu^{2+}$ at $A$-sites ($t_{2g}$ Jahn-Teller effect). As discussed, already in the electronic models for $CuCr_2S_4$ by Lotgering (claiming non-magnetic $Cu^+$) or by Goodenough ($Cu^{2+}$, yet with completely delocalized $3d$ electrons), this is very unlikely. Other candidates could be the Cr ions. In the octahedral $B$-sites, $Cr^{4+}$ and $Cr^{3+}$ could be $t_{2g}$ Jahn-Teller active and may cause dynamic distortions affecting the magnetic structure.

In $CuCr_2S_4$ and $CuCr_2Se_4$, the transition around $T_2$ observed in magnetization, and μSR has an apparent relation to changes in resistivity that have been associated with charge order among $Cr^{3+}$ and $Cr^{4+}$ [13, 15] below $T_2$, while at higher temperatures mixed valence is realized. At high temperatures, the spin structure is collinear ferromagnetic. At low temperatures, our results suggest that the spin structure becomes non-collinear. In fact, recent neutron diffraction studies of our samples reveal clear deviations from a collinear ferromagnetic structure below 80 K and evidence of a conical arrangement of spins at low temperatures. In addition, an analysis of the magnetization data on $CuCr_2Se_4$ single crystals provided evidence for a spin reorientation at low temperatures that supports the results of our μSR studies. These results will be published elsewhere.

## 6. Conclusiones

Our μSR studies, together with magnetization measurements on the two spinel compounds $CuCr_2S_4$ and $CuCr_2Se_4$, reveal magnetic transitions far below the known ferromagnetic ordering temperatures of these metallic systems. The transitions at 80 K and 60 K, respectively, are remarkably stronger visible in the μSR signals than in magnetization due to relative changes in contact and dipolar fields at the muon sites upon changes in magnetic structure. The related changes in electrical resistivity suggest that there is a transition from the mixed valence of $Cr^{3+}$ and $Cr^{4+}$ at high temperatures with collinear ferromagnetism to a charge-ordered state at low temperatures with a conical or even more complex structure as proposed previously for $CuCr_2Se_4$ [15]. Further anomalies around 30 K and 15 K, respectively, are less expressed in changes in ZFC and FC magnetization and may be associated with changes in anisotropy.

The low temperature magnetic transitions found here have strong similarities with results on the isostructural $Fe_{1-x}Cu_xCr_2S_4$ system and suggest a possible relation to Jahn-Teller lattice instabilities in the Cr system.

## Acknowledgment

E.S. acknowledges funding within the MARIA REICHE POSTDOCTORAL FELLOWSHIPS of TU Dresden financed by the Freestate of Saxony and the Federal and State Program for Women Professors (Professorinnenprogramm). Further support came in part from the Deutsche Forschungsgemeinschaft through SFB 1143 (project-id 247310070), SFB 1415 (project-id ID: 417590517), and the Würzburg-Dresden Cluster of Excellence on Complexity and Topology in Quantum Matter-ct.qmat (EXC 2147, project-id 390858490). L.P. and V.T. acknowledge the support from the project ANCD 20.80009.5007.19 (Moldova).

## Appendix

Fig. 10 presents the field dependence of the magnetization at different temperatures in $CuCr_2S_4$ and $CuCr_2Se_4$.

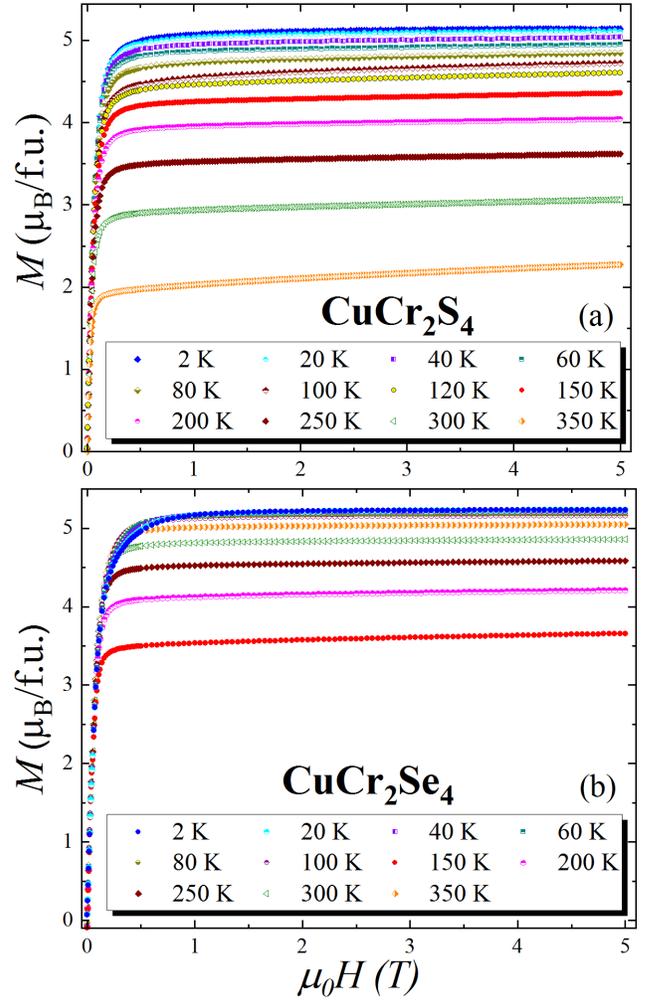

Figure 10: Field dependence of the magnetization at different temperatures for $CuCr_2S_4$ (a) and $CuCr_2Se_4$ (b).